# Cosmic Rays and the Askaryan Effect Reveal Subsurface Structure and Buried Ice on the Moon

E. S. Costello[1,2] , R. R. Ghent[3] , A. Romero-Wolf[4] , P. W. Gorham[5], P. G. Lucey[1,2] , C. J. Tai Udovicic[1,2], P. Linton[6], A. Ludwig[4] , K. McBride[7] , C. Miki[5] , E. Oberla[5] , J. Rolla[4] , and A. Jung[5]

[1]Department of Earth and Planetary Science, University of Hawai'i at Mānoa, Honolulu, HI, USA, [2]Hawaii Institute of Geophysics and Planetology, University of Hawai'i at Mānoa, Honolulu, HI, USA, [3]Planetary Science Institute, Tuscon, AZ, USA, [4]Jet Propulsion Laboratory, California Institute of Technology, Pasadena, CA, USA, [5]Department of Physics, University of Hawai'i at Mānoa, Honolulu, HI, USA, [6]University of Ohio, Athens, OH, USA, [7]Department of Astronomy and Astrophysics, University of Chicago, Chicago, IL, USA

**Abstract** We present the first full-wavelength numerical simulations of the electric field generated by cosmic ray impacts into the Moon. Billions of cosmic rays fall onto the Moon every year. Ultra-high energy cosmic ray impacts produce secondary particle cascades within the regolith and subsequent coherent, wide-bandwidth, linearly-polarized radio pulses by the Askaryan Effect. Observations of the cosmic ray particle shower radio emissions can reveal subsurface structure on the Moon and enable the broad and deep prospecting necessary to confirm or refute the existence of polar ice deposits. Our simulations show that the radio emissions and reflections could reveal ice layers as thin as 10 cm and buried under regolith as deep as 9 m. The Askaryan Effect presents a novel and untapped opportunity for characterizing buried lunar ice at unprecedented depths and spatial scales.

**Plain Language Summary** Particles traveling at extreme speeds, or cosmic rays, impact the Moon so fast that they cause particle cascades that exceed the phase velocity of light (the speed at which a specific point on a light wave travels) in the lunar soil. In a process analogous to a sonic boom, they generate cones of electromagnetic shock waves, the radio emission of which is called the Askaryan Effect. The Askaryan phenomenon produces uniquely identifiable radio waves and retains information about subsurface layers the way radar or sonar sounding would. Unlike radar, which requires high-power antennas to emit radio signals, sensors listening for the Askaryan Effect hear echoes from naturally occurring cosmic ray impacts, drastically lowering the cost and complexity of these detectors. Here, we show for the first time that the Askaryan Effect would produce observable signals from buried ice layers at the lunar poles, providing a novel and untapped method of probing for buried water ice deposits on the Moon.

## 1. Introduction

Cosmic rays continuously collide with the Earth and Moon, whirled into isotropy by galactic magnetic fields between here and their origin at distant high-energy cosmic events such as collapsing stars and jetting black holes (e.g., Hillas, 1984). These high-energy cosmic events produce a spectrum of ultrahigh-energy cosmic rays (UHECR) with energies from GeV ($10^9$ eV) up to ZeV ($10^{21}$ eV). These cosmic rays collide with nuclei and generate relativistic cascades of secondary charged particles, resulting in strong, coherent, wide-bandwidth, linearly-polarized radio pulses by the Askaryan Effect (Askaryan, 1962; Jelley et al., 1965) in a process physically analogous to a sonic boom. A sonic boom occurs when an object travels faster than the speed of sound and generates a cone of constructively interfering waves of air pressure; A cosmic-ray particle cascade travels faster than the phase velocity of light in a dielectrically dense media and generates a cone of coherent radiation (Cherenkov radiation) (James, 2022). Similar to the conical shock wave generated by a sonic boom that extends outward and rearward, the conical radio Cherenkov field extends outward and rearward relative to the cosmic ray and its path through the regolith. On an airless body like Earth's Moon, secondary particle cascades within the dense media of the regolith (James, 2022), starting at ∼1 m deep along the track of the incoming cosmic ray and extending up to 10 m in length at the highest energies, produce strong, coherent, wide-bandwidth, linearly-polarized radio pulses (Askaryan, 1962; Jelley et al., 1965).

UHECR-induced pulses by the Askaryan Effect occur across the Solar System anywhere a cosmic ray enters a dense medium, be it the regolith of an airless body or an atmosphere. Such UHECR-induced pulses are routinely observed in terrestrial atmospheric cascades by ground arrays (Schröder, 2017), and have been observed by suborbital payloads from distances up to 700 km or more (Hoover et al., 2010). UHECR-induced pulses must also occur on the Moon when they traverse the regolith (Askaryan, 1962; Jelley et al., 1965). However, no simulations or instrumentation have yet utilized the well-established phenomena of the Askaryan Effect in the geophysical study of the Moon or any other airless body.

We focus on the Moon and present the first full-wavelength numerical simulations of the electric field generated by UHECR-induced pulses within the lunar regolith. A previous theoretical study of the Askaryan effect in various media, including lunar regolith, was limited by scaling relations using a Monte Carlo model of shower development rather than a full-wavelength simulation of the electric field, and did not include an investigation of stratigraphy or geologic diversity (Alvarez-Muniz et al., 2006). Our full-wavelength numerical simulations demonstrate that subsurface structure and anomalies such as buried ice interact in unique ways with UHECR-induced pulses and represent a novel and untapped solution to an open problem facing lunar science and exploration today: prospecting for lunar ice both deeply and widely enough to discover its presence anywhere within the uppermost 40 m.

The Moon's polar regions feature favorable geologic conditions for preserving water ice over geologic time. The permanently shadowed regions (PSRs) at the Moon's poles and the poles of the planet Mercury both maintain the thermal conditions necessary to harbor tens-of-meters thick deposits of nearly pure water ice for geologic time (e.g., Bertone et al., 2023; Landis et al., 2022; Paige et al., 1992). Remote and Earth-based radar and multispectral investigations have unambiguously discovered ice within the PSRs of Mercury (e.g., Butler et al., 1993; Lawrence et al., 2013; Slade et al., 1992). However, observations of the Moon's PSRs have not discovered the same. Earth-based radar, which would reveal ice within the upper 5 m, has found no unambiguous ice (B. A. Campbell & Campbell, 2006; Stacy et al., 1997).

The Moon's PSRs are not ice-rich like Mercury's, but they are not completely dry. The most unambiguous discovery of water within the PSRs of the Moon comes from the Lunar Crater Observation and Sensing Satellite (LCROSS) impact, which excavated a plume rich in water and other volatiles when it struck the permanent shadow of Cabeus crater (Colaprete et al., 2010); however, the plume indicated a mixture of water and regolith from an excavation zone only about 6 m deep (Luchsinger et al., 2021). Spectral signatures of ice at the optical surface of lunar PSRs have been discovered (Fisher et al., 2017; Hayne et al., 2015; Li et al., 2018); however, the surface ice is patchily distributed (only about 10% of all PSR area) and less than a millimeter thick. Compare this to the ice on Mercury, which covers all available PSR areas with deposits of almost pure water ice thought to be tens of meters deep (Deutsch et al., 2018).

Perhaps Mercury-like ice deposits exist on the Moon but have been buried under a regolith layer (Figure 1). Craters at the south pole of the Moon are statistically shallower than craters at lower latitudes, perhaps because they have been filled with ice and a thick layer of regolith (Rubanenko et al., 2019). Neutron spectroscopy and radar can help us understand how deep such a layer must be. Epithermal neutron fluxes, like those sensed by the Lunar Prospector mission, can be used to determine the abundance of hydrogen within the top meter, and data suggest that the poles are indeed more hydrated than lower latitudes and upper meter of PSRs like Cabeus and Shoemaker may have up to 10% water ice, but that substantial and pure ice deposits do not reside within the top meter (Sanin et al., 2012, 2017). Earth-based delay-Doppler radar observations, which sense the uppermost 5 m or so, lack the clear, coherent backscatter signatures exhibited in the Mercury polar data and show polarization and backscatter patterns and properties that can be explained by roughness and volumetric scattering from rocks rather than ice (B. A. Campbell et al., 2003; D. B. Campbell et al., 2006; Spudis et al., 2013). Thus, Mercury-like ice on the Moon does not reside within the uppermost 5 m.

Models assist us in understanding the possible distribution of deeply buried ice on the Moon but tell a complicated story of unevenly distributed and impact-mixed ice and regolith (Figure 1). Models of ice delivery by ancient large hydrated impacts show that thick "gigaton" ice deposits may be buried and preserved many meters below thick crater ejecta (Cannon et al., 2020); however, such deposits may also have been partially disrupted by ballistic sedimentation (Tai Udovicic et al., 2023). Models of impact gardening suggest that small impacts over billions of years may be randomly excavating deeply buried ice deposits (E. S. Costello et al., 2020, 2021); thus, observations of patchy surface water may indicate more substantial ice deposits buried below. In addition, patchy surface ice may be the result of the migration of water over the last several thousand years rather than evidence of subsurface deposits (Hurley et al., 2023; Schorghofer & Williams, 2020). Ice may also accumulate in thin, centimeters to decimeters-thick layers, sourced by smaller impacts, solar wind, and water molecules which hop on ballistic trajectories until they stick within a PSR (e.g., Prem et al., 2015; Schorghofer, 2015). The resulting ice deposits are likely relatively thin compared to the 'gigaton' deposits that may result from large ancient impacts, only centimeters to decimeters thick. Water that accumulates at the surface in micro cold traps is disturbed by impacts over short geologic timescales (<2 k years; E. Costello & Lucey, 2024; Farrell et al., 2019). The depth of the intensely impact-mixed zone can help us understand limits on the depth to pure ice (E. S. Costello et al., 2021). The surface is mobile, but with increasing depth, it becomes more probable that pure ice could be preserved. It is possible that even thin, centimeters to decimeters-thick deposits have been stochastically both excavated and sequestered by impact gardening below the regolith. The top 5 m are patchily mixed over the last billion years and completely pulverized over 3 billion years (E. S. Costello et al., 2021); thus, we must look deep on the scale of >5 m if we wish to discover large deposits of ice.

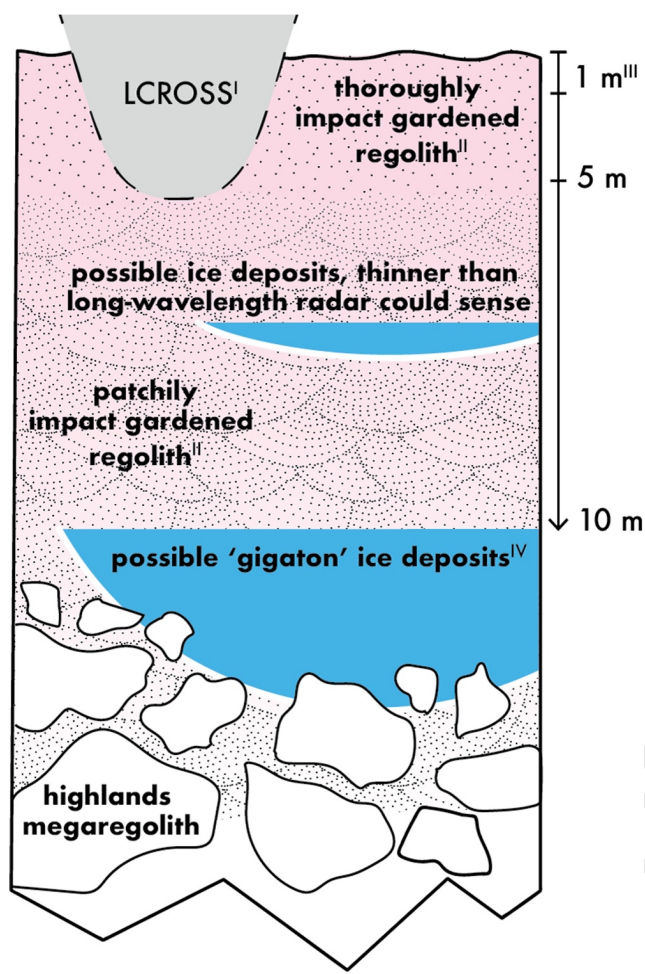

**Figure 1.** Models and observations provide expectations for the depth and survival of lunar ice. (a). Approximately 5 m deep excavation by LCROSS (but no Mercury-like ice) (Colaprete et al., 2010; Luchsinger et al., 2021). (b). Impact gardening thoroughly mixes 1 m and disturbs 5 m over 1 Ga (E. Costello & Lucey, 2024). (c). Neutron spectra suggest Hydrogen is enhanced in the upper meter of regolith in the South Polar region, and within the upper meter of permanently shadowed regions (PSRs), but not enough to indicate the presence of >10% water ice (Lawrence et al., 2013; Sanin et al., 2017). (d). Ancient impacts and volcanism may have filled PSRs with deep, thick "gigaton" ice deposits that have been subsequently buried under meters of regolith (Cannon et al., 2020).

Changes in illumination conditions due to the evolution of the Moon's spin axis may mean that the spatial distribution of PSRs in the past is not the same today (Schörghofer & Rufu, 2023), shrinking the area of the present-day PSRs that could have captured and preserved ancient (>1 Ga) Mercury-like ice deposits on the Moon. Thus, it can be concluded that the spatial distribution of possible large-scale ice deposits on the Moon is uncertain, both vertically and horizontally. Prospecting for ice only where it is cold or where there is surface frost is insufficient.

If Mercury-like ice deposits exist on the Moon, as established by previous radar experiments and discussed above, they must lie below the depth to be probed by the Volatiles Investigating Polar Exploration Rover (VIPER) drill, below the sensing depth of neutron spectrometers, and below the depth that can be reliably sensed by traditional orbital and Earth-based radar. Ice may also be distributed in unexpected places, which reflects the complicated evolution of PSRs on the Moon by the evolving spin axis and stochastic excavation and burial by impacts. To conclusively discover or refute the existence of buried ice deposits on the Moon, a technique that allows for broad prospecting of the deep subsurface (>5 m) with sufficient sensitivity to discover ice deposits that are possibly less than a meter thick is necessary.

In this paper, we present the results of simulations of the detection of deeply buried ice layers. The Askaryan effect emission process has been demonstrated in numerous laboratory measurements over the last two decades (P. W. Gorham et al., 2004, 2007; Saltzberg et al., 2001), and we apply it now to the search for ice on the Moon.

## 2. Simulations: Methods and Results

To investigate the nature of the signals generated by cosmic ray impacts and the Askaryan Effect, we have conducted a series of finite-difference time domain (FDTD) numerical simulations using the program Remcom's XFdtd®. The models are validated by comparisons with direct measurements of accelerator-based particle cascades in dielectric materials measured at particle accelerators, for example, in silica sand (Saltzberg et al., 2001), rock salt (P. W. Gorham et al., 2005), and solid ice (P. W. Gorham et al., 2007), and by semi-analytical models based on extensively validated Monte Carlo codes for shower electrodynamics (Alvarez-Muniz et al., 2011). The simulation space is shown in Figure 2 and consists of a 50 × 10 × 5 m box filled with lunar regolith. The regolith is layered, with increasing layer thickness and increasing density with depth. We use the relationship

$$\epsilon_i = 1.93^{\rho_i} \tag{1}$$

where $\epsilon$ and $\rho$ are the relative permittivity (dielectric constant) and density, respectively, of layer $i$ (Olhoeft & Strangway, 1975). We use absorbing boundary conditions around the box to prevent internal reflections. We place a reflective layer—here, an ice layer—at some depth in the regolith stack. Both the regolith/space interface and the top and bottom surfaces of the reflective layer have fractal roughness. We introduce a current to represent the current created by the particle shower, which XFdtd uses to apply the electromagnetic simulation. We approximate the current using a sequential array of source elements with the appropriate geometry and sequential delays and amplitudes to approximate the propagation speed ($\sim c$) and charge profile of the shower charge bunch. With an appropriate current source, The FDTD method ensures that Maxwell's equations are accurately applied to the dielectric materials present, and the XFdtd simulation engine models the propagation of the EM fields through the simulation space. We record the components of the electric field at an array of point sensors on and above the surface and a planar sensor bisecting the simulation space (Figure 2). These virtual sensors allow us to investigate the signal detectable by a single surface or orbitally deployed sensor at various geometries.

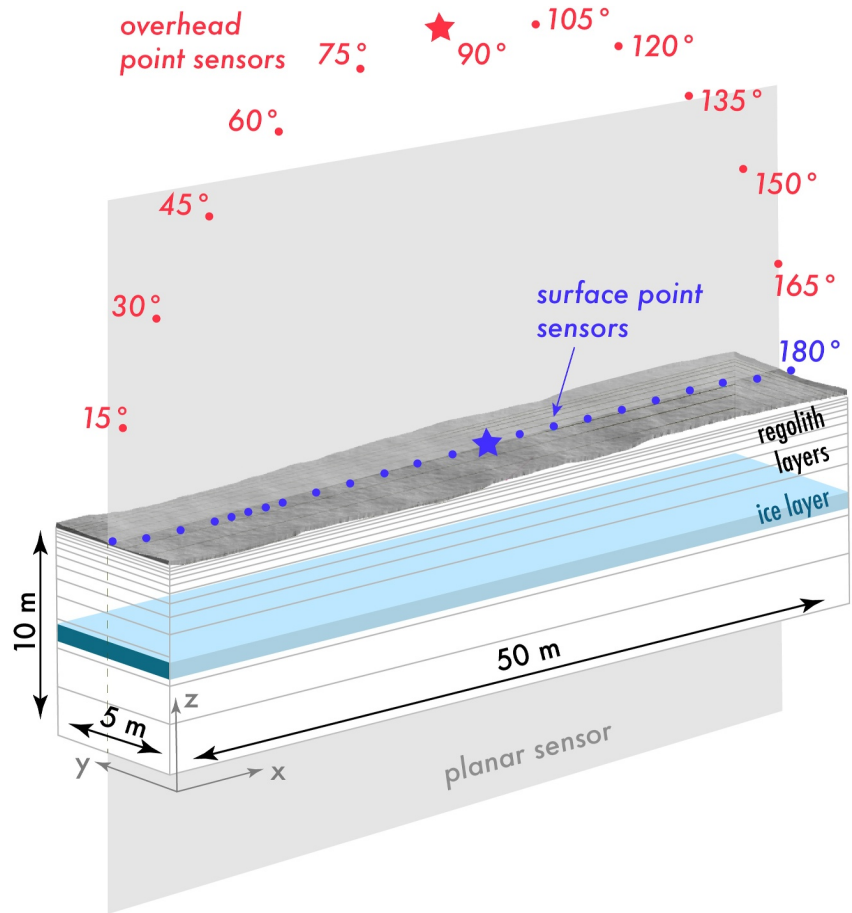

**Figure 2.** A schematic cross-section showing the model setup. We collect E-field strength from a planar sensor, which bisects the longitudinal plane of the model (shaded gray), as well as overhead (red) and surface point (blue) sensors. The planar sensor captures the full radio wave propagation. Stars indicate the point sensors for which we generate the point-sensor plots, which are shown in subsequent figures, and represent the point measurements that would be observed by an instrument in orbit or on the surface. Here, the regolith layers are made invisible to illustrate our treatment of layering within the column of the regolith above the ice layer.

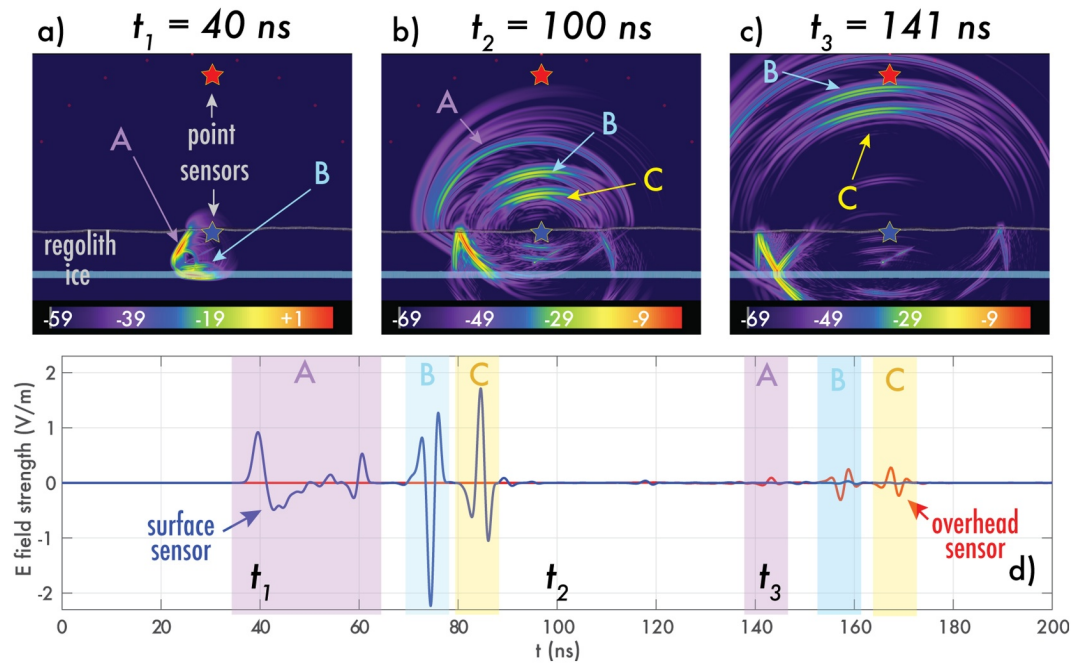

**Figure 3.** Results of a numerical simulation of Askaryan-induced radio wave propagation through regolith (transparent in this figure) with an embedded ice layer 1 m thick buried at 6 m depth. This simulation run is for a $10^{19}$ eV cosmic ray (10 EeV). (a–c) show the total electric field in dB relative to 10 V/m reference value (colors) as recorded by the planar sensor shown in Figure 2 at 40, 100, and 141 ns, respectively. The wavefront labeled A is the direct emission. The wavefront labeled B is the reflected signal from the top of the ice layer. The wavefront labeled C is the reflection from the base of the ice layer. (d) A plot showing the total electric field versus time at the surface sensor located at the center of the model (blue curve) and at the overhead sensor directly above (red curve). Colored boxes labeled A, B, and C correspond to the direct and reflected wavefronts above.

Here, we show the results of one model run simulating the propagation of EM waves from a particle shower caused by the oblique impact of a $10^{19}$ eV cosmic ray. This model has a 1 m thick ice layer embedded in the regolith at a depth of 6 m in detail from overhead and surface sensors (Figure 3) and from a variety of overhead angles (Figure 4). Four further simulations, with variations in ice layer thickness and depth, demonstrate that the effects of a 1 cm thick ice layer and an ice layer buried under 9 m of regolith are detectable (Figure 5.)

We estimate a permittivity of $\epsilon_{ice} \simeq 2.2$, which accounts for the lunar surface gravity and expected overburden compression of an initial low-density ice deposit. Pure, full-density terrestrial Antarctic ice at depths of >100 m has a permittivity of ∼3.2 due to compression by the weight of the overlying ice. However, lunar ice deposits are very unlikely to compress to a comparable density or permittivity within the first 10–20 m of regolith depth in lunar gravity. For example, Brouet et al. (2016) prepared laboratory porous ice samples specifically to investigate their nature in solar system icy bodies and found an average density of ∼400 kg m$^{-3}$ in their samples prepared at 1 g, similar to surface ice in Antarctica. On Earth, South Polar ice compresses to ∼600 kg m$^{-3}$ (corresponding to $\epsilon_{ice} \simeq 2.2$) at depths of 35 m (van den Broeke, 2008), where the cryostatic pressure is 178 kN m$^{-2}$. In lunar gravity, at an average regolith density of 1,800 kg m$^{-3}$, similar pressure requires a depth of 60 m; thus, our permittivity estimate is very conservative and could be significantly lower, which would enhance the reflected signals.

As the cosmic-ray shower develops, it gives rise to a conical EM wavefront that propagates through the simulation space at the Cherenkov angle (e.g., Zas et al., 1992), given by $\cos\theta_C = [\sqrt{\epsilon}\beta]^{-1}$, for shower charge propagating at speed $\beta = v/c \simeq 1$. The emission from the CR shower has a very broad radio spectrum, which includes detectable frequencies from several tens of MHz up to ∼1GHz or more. At ≤100 MHz frequencies corresponding to wavelengths comparable to the several-meter-long shower, the emission becomes much broader due to diffraction effects. The direct signals radiated away from the primary Cherenkov direction arise from the lower-frequency portion of the frequency spectrum than higher-frequency reflected signals; this part of the signal is shown in the cross section in Figures 3a–3c as the "upper branch" of the conical wavefront. This branch propagates upward, contacting the regolith-space boundary after a few ns; it is then refracted and continues to propagate upward into space. When it reaches one of our simulated sensors, we refer to that portion of the signal as the "direct emission."

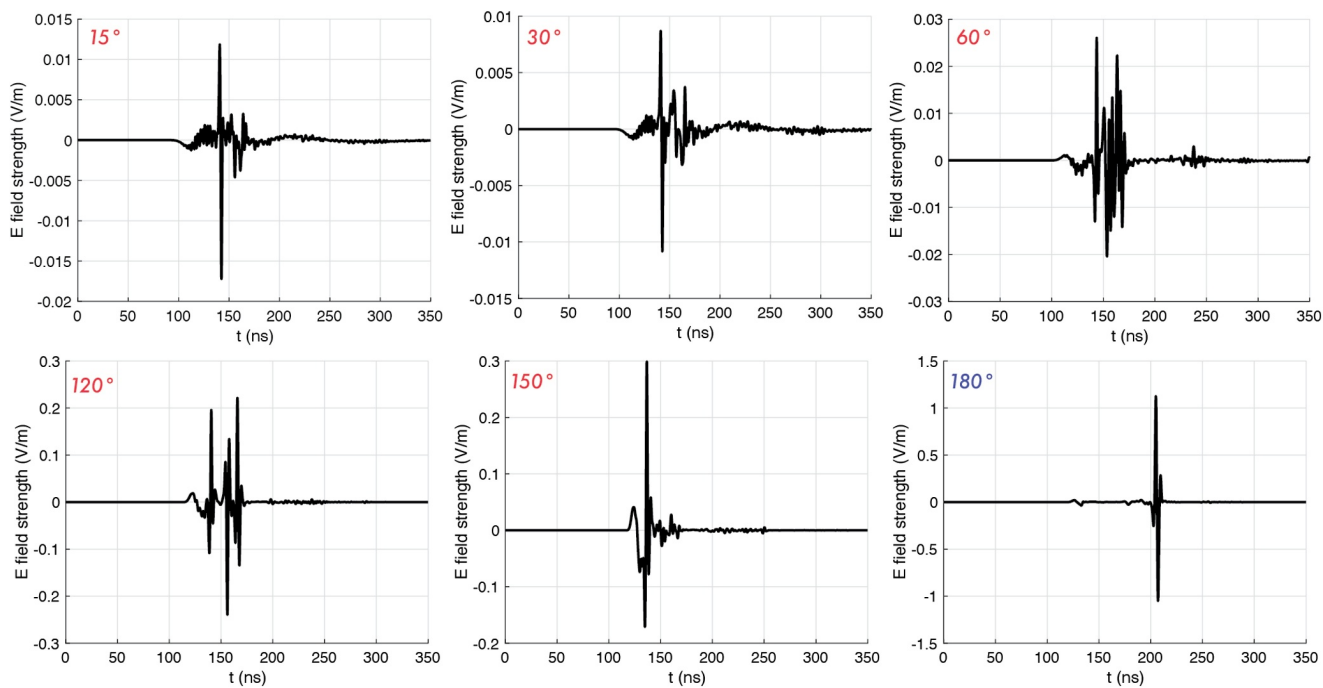

**Figure 4.** Results of the numerical simulation of Askaryan-induced radio wave propagation at six additional point sensors demonstrate that a sensor placed at any of these six positions would detect the signal generated by the shower shown in Figure 3. Degrees markers correspond to those point sensors shown in Figure 2.

The higher-frequency emission is more forward-directed; thus, both the angular and frequency spectra cause preferential intensity weighting for the forward signals. This forward signal is shown in Figures 3a–3c as the "lower branch" of the propagating EM wave, which contacts the 6 m deep ice layer at 40 ns. Some of that energy is reflected from the top surface of the ice layer, propagating upward, eventually reaching the surface and overhead sensors at later times. The rest of the energy propagates through the ice layer, and some reflects from the bottom ice/regolith interface.

While the most favorable geometry is more constrained to a cone around the Cherenkov angle for direct signals, Figure 4 shows that a signal is observable at various observation geometries, and the reflected signal is prominent for a wider range of geometries, peaking closer to zenith even for an inclined shower (see Figure 3).

The plot in Figure 3d shows the total electric field strength as a function of time following shower initiation; the peaks in these plots corresponding to the arrival of various wave fronts are labeled. The signal strength at the overhead sensor is smaller than the surface sensor because the decrease in signal strength follows a 1/R law versus distance. In addition, sensors at the surface do not suffer losses from the Fresnel reflection moving from the surface to vacuum.

Figures 3b and c show that both the top and bottom ice surfaces produce reflections with total field strength larger than, or comparable to, the strength of the direct emission. Figures 3b and c also show that the polarity of the signal is reversed at the top and bottom of the ice layer, due to the fact that the ice has a lower permittivity than the surrounding regolith. Moreover, for sufficiently thick ice layers, it is possible to distinguish reflections from the top and bottom ice surfaces.

## 3. Discussion and Conclusions

Over 10 billion 1 EeV and 10 million 10 EeV cosmic rays like those we model here strike the Moon each year (from Equation 9 in Abreu et al. (2021)) scaled to the surface area of the Moon). We show that a single point sensor deployed in or near a PSR or in orbit could sense the electromagnetic response of buried ice, and a surface sensor may be able to detect the signals generated by exponentially more numerous lower-energy cosmic ray impacts due to higher signal strength at the surface. Each year, over twenty-six thousand 1 EeV cosmic rays fall into the PSR of Cabaeus crater alone, and six thousand 10 EeV cosmic rays fall into south-polar PSRs (PSR area from Mazarico et al. (2011)). At energies $\ll 1$, exponentially more cosmic rays fall every year, generating signals which could be detectable by orbital and surface sensors. Because of proximity to the event and because it does not suffer loss due to Fresnel reflection moving from the surface to vacuum, a surface detector has approximately a three orders of magnitude lower detection threshold ($\approx 10^{15}$ eV = 1 PeV) compared to an orbital detector (Figure 3d). For a surface instrument, compared to an orbital instrument, the detectable cosmic ray rate increases as the energy threshold to the power $-1.6$ to about 170 events per day $((0.001)^{-1.6} \approx 60,000$ times the rate at 1 EeV).

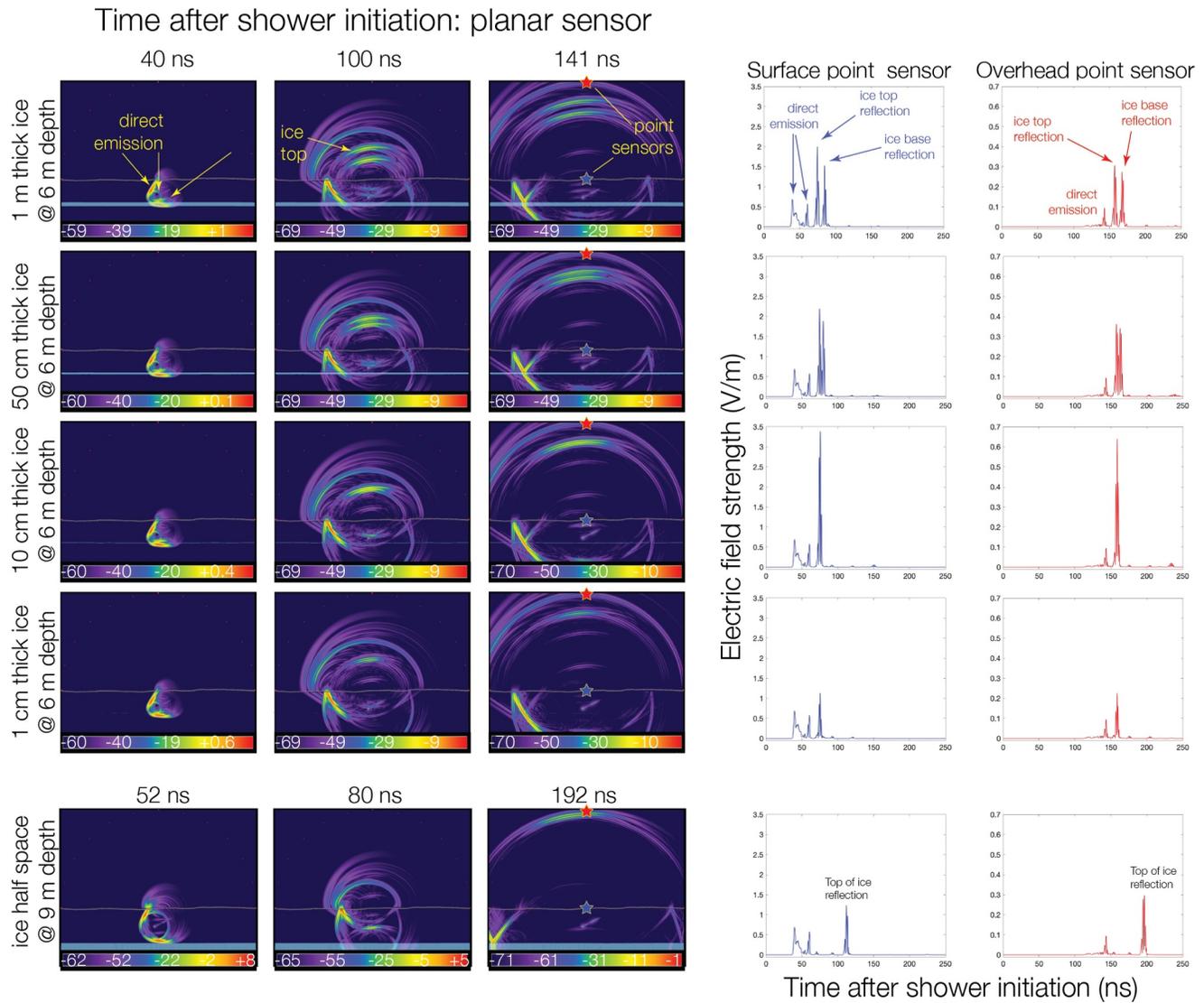

**Figure 5.** Results of five numerical simulations of Askaryan-induced radio wave propagation through regolith (transparent in this figure) with embedded ice layers of varying thickness and depth. Left three columns: total electric field in dB relative to 10 V/m reference value (colors) as recorded by the planar sensor shown in Figure 2 at the time steps indicated. View is down the *Y*-axis. The right two columns show plots of the total electric field versus time at the surface sensor located at the center of the model, and at the overhead sensor directly above. Peaks associated with the direct emission and reflections from the ice layer are labeled with colored boxes in the top and bottom rows. These simulation runs are for a $10^{19}$ eV cosmic ray.

The sensor instrument could be viewed as a passive radar system. The radar system would be bistatic, with a subsurface "transmitter" (the cosmic ray shower) viewed from a receiver system on the surface or in low lunar orbit. Such an instrument would benefit from low-power passive sensing rather than high-power active sounding, showing promise for this novel, next-generation technique.

Recent studies have investigated if a passive radio receiver could be sensitive over a bandwidth of 150–800 MHz (Bramson et al., 2023; Prechelt et al., 2022), similar in magnitude to the UHECR Askaryan shower on the Moon that we model here. Initial results are promising. Wide and deep prospecting is a necessity, given the likely heterogeneous and deeply buried nature of lunar ice deposits. Advantages of a remote sensing technique that takes advantage of UHECR-induced pulses include the ability to probe tens of meters deep into the lunar subsurface, like GPR, but over the entire surface of the Moon - an orders-of-magnitude improvement on the ∼1km range limitation imposed by a rover-mounted GPR. The cosmic ray particle shower signal source is similar to a buried dipole antenna, so remote sensing would not suffer the loss of signal or clutter associated with a strong surface reflection like traditional active remote radar (Hayne et al., 2014, e.g., Section 4.4.2). Further, the simulations presented here show that this technique is sensitive to ice layers that would be too thin to be detected by traditional long-wavelength sounding or synthetic aperture radar.

The potential of this phenomenon to illuminate lunar ice deposits warrants additional quantitative analysis, including the reconstruction of cosmic ray direction and strength from a signal and the localization of the emission source from orbit. Both reconstruction and localization are the subject of ongoing simulation work. Analysis should be tailored to specific orbital or landed mission opportunities; for example, an orbital mission will require further analysis of altitude constraints on detectability, and a landed mission must focus on detection rates. A single deployed surface sensor will observe only within a limited detection radius but could trade broad prospecting for sensitivity to signals produced by a broader range of cosmic ray energies (e.g., exponentially more abundant cosmic rays at energies $\ll 1$ EeV), thus exponentially improving the local signal detection rate. Here, we have explored feasibility through demonstration of the phenomena; however, robust modeling of detection rates for specific mission scenarios will be necessary to constrain the feasibility of specific orbital or landed missions. Further models will also assist in constraining signal-to-noise; however, the frequency spectrum, coherence, and polarization of Askaryan emission are unique compared to incoherent thermal noise and radio noise from cosmic sources. Balloon-borne payload measurements of radio pulses from analogous processes on Earth in Antarctica show Cherenkov-like signals can be distinguished even in the presence of strong anthropogenic interference (P. Gorham et al., 2021, e.g.).

In the future, we should also expand full-wavelength electromagnetic simulations to set constraints on the detectability of ice at various extreme depths and thicknesses. Figure 5 shows a limited sample of possibilities, but we could also use a similar simulation setup to analyze the effects of ice-regolith mixtures and scattering by subsurface rocks. Further, although Cherenkov emission typically develops in the upper 1–5 m, it may still be useful for detecting shallow or regolith-ice mixtures, and the possibility merits further investigation. Highly inclined cosmic rays emit within the first meter of regolith and may, therefore, be particularly worthy of further analysis in the context of investigating shallow ice.

We have focused on the tantalizing potential of the Askaryan Effect to reveal ice on the Moon; however, these simulations also demonstrate the broader potential of examining Askaryan-induced emissions on planetary bodies. For example, our simulations illustrate the potential for using Askaryan-induced emissions to broadly probe the structure of the lunar subsurface over a wide range of expected regolith depths, from maria to highlands. In a manner similar to that illustrated above, reflections can be simulated from a variety of useful subsurface scenarios, any of which we could expect to encounter during a single lunar orbit, including deep highland regolith, shallow mare regolith over mare basalt, thin or thick layers of pure ice in the polar regions, or layers of impure ice or icy regolith. Beyond the Moon, Askaryan-induced emissions may reveal subsurface layering or lakes on icy satellites, such as the astrobiologically significant Europa. Further simulations like those we present here can qualify and expand on the broadly impactful possibilities for the fundamental phenomena, as-yet underutilized in planetary science: The Askaryan Effect.

## Data Availability Statement

The figures shown in this manuscript will be archived on Figshare with cite-able DOI at the time of acceptance. The software, Remcom's XFdtd®, is documented on the Remcom website (https://www.remcom.com/xfdtd-3d-em-simulation-software). XFdtd code, Matlab code, and tables of outputs necessary to reproduce the figures and results shown in this manuscript are included in full as supplemental material for review and will be archived with citable DOI on Zenodo with cite-able DOI upon manuscript acceptance.

**Acknowledgments**
This work is supported by NASA Development and Advancement of Lunar Instrumentation Grant 80NSSCK0851.

## References

Abreu, P., Aglietta, M., Albury, J. M., Allekotte, I., Almela, A., & Alvarez-Muñiz, J. (2021). The energy spectrum of cosmic rays beyond the turn-down around 10 17 10 17 ev as measured with the surface detector of the pierre auger observatory. *The European Physical Journal C*, *81*, 1–25.

Alvarez-Muniz, J., Marqués, E., Vázquez, R. A., & Zas, E. (2006). Coherent radio pulses from showers in different media: A unified parametrization. *Physical Review D - Particles, Fields, Gravitation and Cosmology*, *74*(2), 023007. https://doi.org/10.1103/physrevd.74.023007

Alvarez-Muniz, J., Romero-Wolf, A., & Zas, E. (2011). Practical and accurate calculations of askaryan radiation. *Physical Review D - Particles, Fields, Gravitation and Cosmology*, *84*(10), 103003. https://doi.org/10.1103/physrevd.84.103003

Askaryan, G. A. (1962). Excess negative charge of an electron-photon shower and its coherent radio emission. *Soviet Physics Journal of Experimental and Theoretical Physics*, *14*(2), 441–443.

Bertone, S., Mazarico, E., Barker, M. K., Siegler, M. A., Martinez-Camacho, J. M., Hamill, C. D., et al. (2023). Highly resolved topography and illumination at mercury's south pole from messenger mdis nac. *The Planetary Science Journal*, *4*(2), 21. https://doi.org/10.3847/psj/acaddb

Bramson, A., Gorham, P., Allison, P., Andrew, M., Bailey, S., & Beatty, J. (2023). Cryptex: A mission concept to test the presence, properties, and geophysical context of lunar cryptomaria. *LPI Contribution*, *2806*, 1797.

Brouet, Y., Neves, L., Sabouroux, P., Levasseur-Regourd, A. C., Poch, O., Encrenaz, P., et al. (2016). Characterization of the permittivity of controlled porous water ice-dust mixtures to support the radar exploration of icy bodies. *Journal of Geophysical Research: Planets*, *121*(12), 2426–2443. https://doi.org/10.1002/2016je005045

Butler, B. J., Muhleman, D. O., & Slade, M. A. (1993). Mercury: Full-disk radar images and the detection and stability of ice at the north pole. *Journal of Geophysical Research*, *98*(E8), 15003–15023. https://doi.org/10.1029/93je01581

Campbell, B. A., & Campbell, D. B. (2006). Regolith properties in the south polar region of the moon from 70-cm radar polarimetry. *Icarus*, *180*(1), 1–7. https://doi.org/10.1016/j.icarus.2005.08.018

Campbell, B. A., Campbell, D. B., Chandler, J. F., Hine, A. A., Nolan, M. C., & Perillat, P. J. (2003). Radar imaging of the lunar poles. *Nature*, *426*(6963), 137–138. https://doi.org/10.1038/426137a

Campbell, D. B., Campbell, B. A., Carter, L. M., Margot, J.-L., & Stacy, N. J. (2006). No evidence for thick deposits of ice at the lunar south pole. *Nature*, *443*(7113), 835–837. https://doi.org/10.1038/nature05167

Cannon, K. M., Deutsch, A. N., Head, J. W., & Britt, D. T. (2020). Stratigraphy of ice and ejecta deposits at the lunar poles. *Geophysical Research Letters*, *47*(21), e2020GL088920. https://doi.org/10.1029/2020gl088920

Colaprete, A., Schultz, P., Heldmann, J., Wooden, D., Shirley, M., Ennico, K., et al. (2010). Detection of water in the lcross ejecta plume. *Science*, *330*(6003), 463–468. https://doi.org/10.1126/science.1186986

Costello, E., & Lucey, P. (2024). The age and evolution of lunar micro cold traps at the scale of surface exploration. *Geophysical Research Letters*, *51*(1), e2023GL105369. https://doi.org/10.1029/2023gl105369

Costello, E. S., Ghent, R. R., Hirabayashi, M., & Lucey, P. G. (2020). Impact gardening as a constraint on the age, source, and evolution of ice on mercury and the moon. *Journal of Geophysical Research: Planets*, *125*(3), e2019JE006172. https://doi.org/10.1029/2019je006172

Costello, E. S., Ghent, R. R., & Lucey, P. G. (2021). Secondary impact burial and excavation gardening on the moon and the depth to ice in permanent shadow. *Journal of Geophysical Research: Planets*, *126*(9), e2021JE006933. https://doi.org/10.1029/2021je006933

Deutsch, A. N., Head, J. W., Chabot, N. L., & Neumann, G. A. (2018). Constraining the thickness of polar ice deposits on mercury using the mercury laser altimeter and small craters in permanently shadowed regions. *Icarus*, *305*, 139–148. https://doi.org/10.1016/j.icarus.2018.01.013

Farrell, W., Hurley, D., Poston, M., Hayne, P., Szalay, J., & McLain, J. (2019). The young age of the lamp-observed frost in lunar polar cold traps. *Geophysical Research Letters*, *46*(15), 8680–8688. https://doi.org/10.1029/2019gl083158

Fisher, E. A., Lucey, P. G., Lemelin, M., Greenhagen, B. T., Siegler, M. A., Mazarico, E., et al. (2017). Evidence for surface water ice in the lunar polar regions using reflectance measurements from the lunar orbiter laser altimeter and temperature measurements from the diviner lunar radiometer experiment. *Icarus*, *292*, 74–85. https://doi.org/10.1016/j.icarus.2017.03.023

Gorham, P., Ludwig, A., Deaconu, C., Cao, P., Allison, P., Banerjee, O., et al. (2021). Unusual near-horizon cosmic-ray-like events observed by anita-iv. *Physical Review Letters*, *126*(7), 071103. https://doi.org/10.1103/physrevlett.126.071103

Gorham, P. W., Barwick, S. W., Beatty, J. J., Besson, D. Z., Binns, W. R., Chen, C., et al. (2007). Observations of the askaryan effect in ice. *Physical Review Letters*, *99*(17), 171101. https://doi.org/10.1103/PhysRevLett.99.171101

Gorham, P. W., Saltzberg, D., Field, R., Guillian, E., Milinčić, R., Miočinović, P., et al. (2005). Accelerator measurements of the askaryan effect in rock salt: A roadmap toward teraton underground neutrino detectors. *Physical Review D - Particles, Fields, Gravitation and Cosmology*, *72*(2), 023002. https://doi.org/10.1103/PhysRevD.72.023002

Gorham, P. W., Saltzberg, D., Field, R. C., Guillian, E., Milincic, R., Walz, D., et al. (2004). Accelerator measurements of the askaryan effect in rock salt: A roadmap toward teraton underground neutrino detectors. *Physics Reviews*, *D72*, 023002. https://doi.org/10.1103/PhysRevD.72.023002

Hayne, P. O., Hendrix, A., Sefton-Nash, E., Siegler, M. A., Lucey, P. G., Retherford, K. D., et al. (2015). Evidence for exposed water ice in the moon's south polar regions from lunar reconnaissance orbiter ultraviolet albedo and temperature measurements. *Icarus*, *255*, 58–69. https://doi.org/10.1016/j.icarus.2015.03.032

Hayne, P. O., Paige, D. A., & Ingersoll, A. P. (2014). New approaches to lunar ice detection and mapping. Keck Institute for Space Studies.

Hillas, A. M. (1984). The origin of ultra-high-energy cosmic rays. *Annual Review of Astronomy and Astrophysics*, *22*(1), 425–444. https://doi.org/10.1146/annurev.astro.22.1.425

Hoover, S., Nam, J., Gorham, P. W., Grashorn, E., Allison, P., Barwick, S. W., et al. (2010). Observation of ultrahigh-energy cosmic rays with the ANITA balloon-borne radio interferometer. *Physical Review Letters*, *105*(15), 151101. https://doi.org/10.1103/PhysRevLett.105.151101

Hurley, D. M., Siegler, M. A., Cahill, J. T., Colaprete, A., Costello, E., Deutsch, A. N., et al. (2023). Surface volatiles on the moon. *Reviews in Mineralogy and Geochemistry*, *89*(1), 787–827. https://doi.org/10.2138/rmg.2023.89.18

James, C. W. (2022). Nature of radio-wave radiation from particle cascades. *Physical Review D*, *105*(2), 023014. https://doi.org/10.1103/physrevd.105.023014

Jelley, J. V., Fruin, J. H., Porter, N. A., Weekes, T. C., Smith, F. G., & Porter, R. A. (1965). Radio pulses from extensive cosmic-ray air showers. *Nature*, *205*(4969), 327–328. https://doi.org/10.1038/205327a0

Landis, M. E., Hayne, P. O., Williams, J.-P., Greenhagen, B. T., & Paige, D. A. (2022). Spatial distribution and thermal diversity of surface volatile cold traps at the lunar poles. *The Planetary Science Journal*, *3*(2), 39. https://doi.org/10.3847/psj/ac4585

Lawrence, D. J., Feldman, W. C., Goldsten, J. O., Maurice, S., Peplowski, P. N., Anderson, B. J., et al. (2013). Evidence for water ice near mercury's north pole from messenger neutron spectrometer measurements. *Science*, *339*(6117), 292–296. https://doi.org/10.1126/science.1229953

Li, S., Lucey, P. G., Milliken, R. E., Hayne, P. O., Fisher, E., Williams, J.-P., et al. (2018). Direct evidence of surface exposed water ice in the lunar polar regions. *Proceedings of the National Academy of Sciences*, *115*(36), 8907–8912. https://doi.org/10.1073/pnas.1802345115

Luchsinger, K. M., Chanover, N. J., & Strycker, P. D. (2021). Water within a permanently shadowed lunar crater: Further lcross modeling and analysis. *Icarus*, *354*, 114089. https://doi.org/10.1016/j.icarus.2020.114089

Mazarico, E., Neumann, G., Smith, D., Zuber, M., & Torrence, M. (2011). Illumination conditions of the lunar polar regions using lola topography. *Icarus*, *211*(2), 1066–1081. https://doi.org/10.1016/j.icarus.2010.10.030

Olhoeft, G. R., & Strangway, D. (1975). Dielectric properties of the first 100 meters of the moon. *Earth and Planetary Science Letters*, *24*(3), 394–404. https://doi.org/10.1016/0012-821x(75)90146-6

Paige, D. A., Wood, S. E., & Vasavada, A. R. (1992). The thermal stability of water ice at the poles of mercury. *Science*, *258*(5082), 643–646. https://doi.org/10.1126/science.258.5082.643

Prechelt, R., Costello, E., Ghent, R., Gorham, P., Lucey, P., Romero-Wolf, A., & Varner, G. (2022). Passive bistatic radar probes of the subsurface on airless bodies using high energy cosmic rays via the askaryan effect. *arXiv preprint arXiv:2212.07689*.

Prem, P., Artemieva, N., Goldstein, D., Varghese, P., & Trafton, L. (2015). Transport of water in a transient impact-generated lunar atmosphere. *Icarus*, *255*, 148–158. https://doi.org/10.1016/j.icarus.2014.10.017

Rubanenko, L., Venkatraman, J., & Paige, D. A. (2019). Thick ice deposits in shallow simple craters on the moon and mercury. *Nature Geoscience*, *12*(8), 597–601. https://doi.org/10.1038/s41561-019-0405-8

Saltzberg, D., Gorham, P., Walz, D., Field, C., Iverson, R., Odian, A., et al. (2001). Observation of the Askaryan effect: Coherent microwave Cherenkov emission from charge asymmetry in high-energy particle cascades. *Physical Review Letters*, *86*(13), 2802–2805. https://doi.org/10.1103/PhysRevLett.86.2802

Sanin, A., Mitrofanov, I., Litvak, M., Bakhtin, B., Bodnarik, J., Boynton, W. V., et al. (2017). Hydrogen distribution in the lunar polar regions. *Icarus*, *283*, 20–30. https://doi.org/10.1016/j.icarus.2016.06.002

Sanin, A., Mitrofanov, I., Litvak, M., Malakhov, A., Boynton, W., Chin, G., et al. (2012). Testing lunar permanently shadowed regions for water ice: Lend results from lro. *Journal of Geophysical Research*, *117*(E12). https://doi.org/10.1029/2011je003971

Schorghofer, N. (2015). Two-dimensional description of surface-bounded exospheres with application to the migration of water molecules on the moon. *Physical Review E*, *91*(5), 052154. https://doi.org/10.1103/physreve.91.052154

Schörghofer, N., & Rufu, R. (2023). Past extent of lunar permanently shadowed areas. *Science Advances*, *9*(37), eadh4302. https://doi.org/10.1126/sciadv.adh4302

Schorghofer, N., & Williams, J.-P. (2020). Mapping of ice storage processes on the moon with time-dependent temperatures. *The Planetary Science Journal*, *1*(3), 54. https://doi.org/10.3847/psj/abb6ff

Schröder, F. G. (2017). Radio detection of cosmic-ray air showers and high-energy neutrinos. *Progress in Particle and Nuclear Physics*, *93*, 1–68. https://doi.org/10.1016/j.ppnp.2016.12.002

Slade, M. A., Butler, B. J., & Muhleman, D. O. (1992). Mercury radar imaging: Evidence for polar ice. *Science*, *258*(5082), 635–640. https://doi.org/10.1126/science.258.5082.635

Spudis, P., Bussey, D., Baloga, S., Cahill, J., Glaze, L., Patterson, G., et al. (2013). Evidence for water ice on the moon: Results for anomalous polar craters from the lro mini-rf imaging radar. *Journal of Geophysical Research: Planets*, *118*(10), 2016–2029. https://doi.org/10.1002/jgre.20156

Stacy, N., Campbell, D., & Ford, P. (1997). Arecibo radar mapping of the lunar poles: A search for ice deposits. *Science*, *276*(5318), 1527–1530. https://doi.org/10.1126/science.276.5318.1527

Tai Udovicic, C., Frizzell, K., Kodikara, G., Kopp, M., Luchsinger, K., Madera, A., et al. (2023). Buried ice deposits in lunar polar cold traps were disrupted by ballistic sedimentation. *Journal of Geophysical Research: Planets*, *128*(5), e2022JE007567. https://doi.org/10.1029/2022je007567

van den Broeke, M. (2008). Depth and density of the antarctic firn layer. *Arctic Antarctic and Alpine Research*, *40*(2), 432–438. https://doi.org/10.1657/1523-0430(07-021)[broeke]2.0.co;2

Zas, E., Halzen, F., & Stanev, T. (1992). Electromagnetic pulses from high-energy showers: Implications for neutrino detection. *Physical Review D*, *45*(1), 362–376. https://doi.org/10.1103/physrevd.45.362